\documentclass{ws-p8-50x6-00}
\begin{document}
\title{Progress in understanding confinement.}
\author{A. Digiacomo}
\address{Universit\`a and I.N.F.N., Via Buonarroti 2, 56100 Pisa\\
E-mail: adriano.digiacomo@df.unipi.it}
\maketitle
\abstracts{
A few aspects of the mechanism of confinement of color
by monopole condensation are reviewed.
}
\section{Introduction.}
In recent years a research program was developed\cite{1,2,3,4} to investigate
dual superconductivity of the vacuum as a mechanism of confinement of color
in QCD. The original idea goes back to t'Hooft\cite{5} and
Mandelstam\cite{6}, and is, modulo duality transformation, the same as
confinement of magnetic charges in a type 2 superconductor.
The end points of an Abrikosov flux tube can be viewed as monopoles;
their energy is proportional to the length of the flux tube, implying
that an infinite energy is needed to pull apart at infinite distance a
monopole antimonopole pair.

However, when going more in detail in the definition of monopoles in non
abelian gauge theories, one discovers that monopoles  are intrinsecally $U(1)$
configurations, so that a suitable $U(1)$ must be defined to expose them. The
procedure is known as ``abelian projection'', and associates a conserved magnetic
charge to any operator in the adjoint representation.

Dual superconductivity can be investigated in connection with any abelian projection.
The technique used is to detect if there is condensation of monopoles in the confined phase, but
not in the deconfined one.

A disorder parameter has been defined\cite{3,4}, as the vacuum expectation value of an operator
carrying magnetic charge. The infinite volume (thermodinamic) limit, where the phase
transition can take place, is reached by a finite size scaling analysis. An
unambigous demonstration emerges that magnetic charges condense in the confined
phase, whilst the Hilbert space is split into superselected sectors of different
magnetic charge in the deconfined regime. This happens in all the abelian projections
which have been investigated, but also in an average over an infinite number of
abelian projections. The conclusion is that the property is independent of 
the specific choice of the abelian projection\cite{4}.
This possibility had already been considered in ref.\cite{7}.

The conclusion is that
the (yet unknown) dual excitations of the theory which condense in the confining
vacuum have non zero magnetic charge in all the abelian projections.

The disorder parameter $\langle\mu\rangle$ has the form
\begin{equation}
\langle\mu\rangle = \frac{\displaystyle \tilde Z}{\displaystyle Z}
\label{eq1}\end{equation}
where $\tilde Z$ is the partition function obtained from $Z$ by introducing in the
action a magnetic defect on the slice of constant time $t$\cite{2,4}.

$\langle\mu\rangle$ is then  roughly an exponential of a volume integral; as such it
fluctuates as te exponential of the square root of the volume, i.e. very wildly. A
way to go around this difficulty was first suggested in ref.\cite{2}: it amounts to
study instead of $\langle\mu\rangle$ the quantity
\[ \rho = \frac{d}{d\beta}\ln \langle\mu\rangle \]
Since $\langle\mu(\beta=0)\rangle = 1$,
\[ \langle\mu\rangle = \exp\int_0^\beta \rho(x) dx\]
The expected shape of $\langle\mu\rangle$ at the phase transition is as 
in fig.1, and
reflects in a shape like in fig.2 for $\rho$.
\par\noindent
\begin{figure}
\begin{minipage}{0.5\textwidth}
{\centerline{
\includegraphics[width=0.9\textwidth]{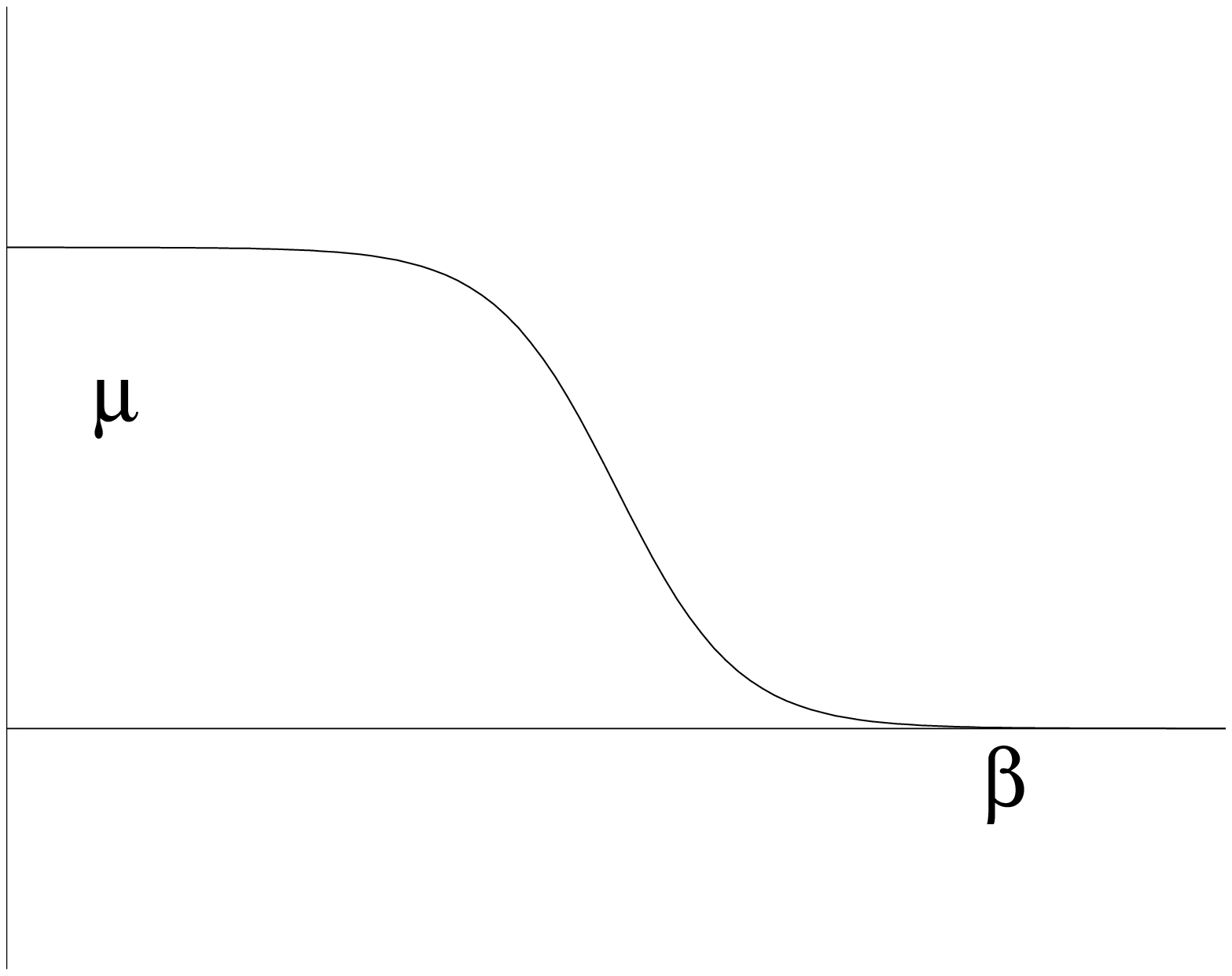}
}}
\caption{$\mu$ vs. $\beta$.}
\end{minipage}
\begin{minipage}{0.5\textwidth}
{\centerline{
\includegraphics[width=0.9\textwidth]{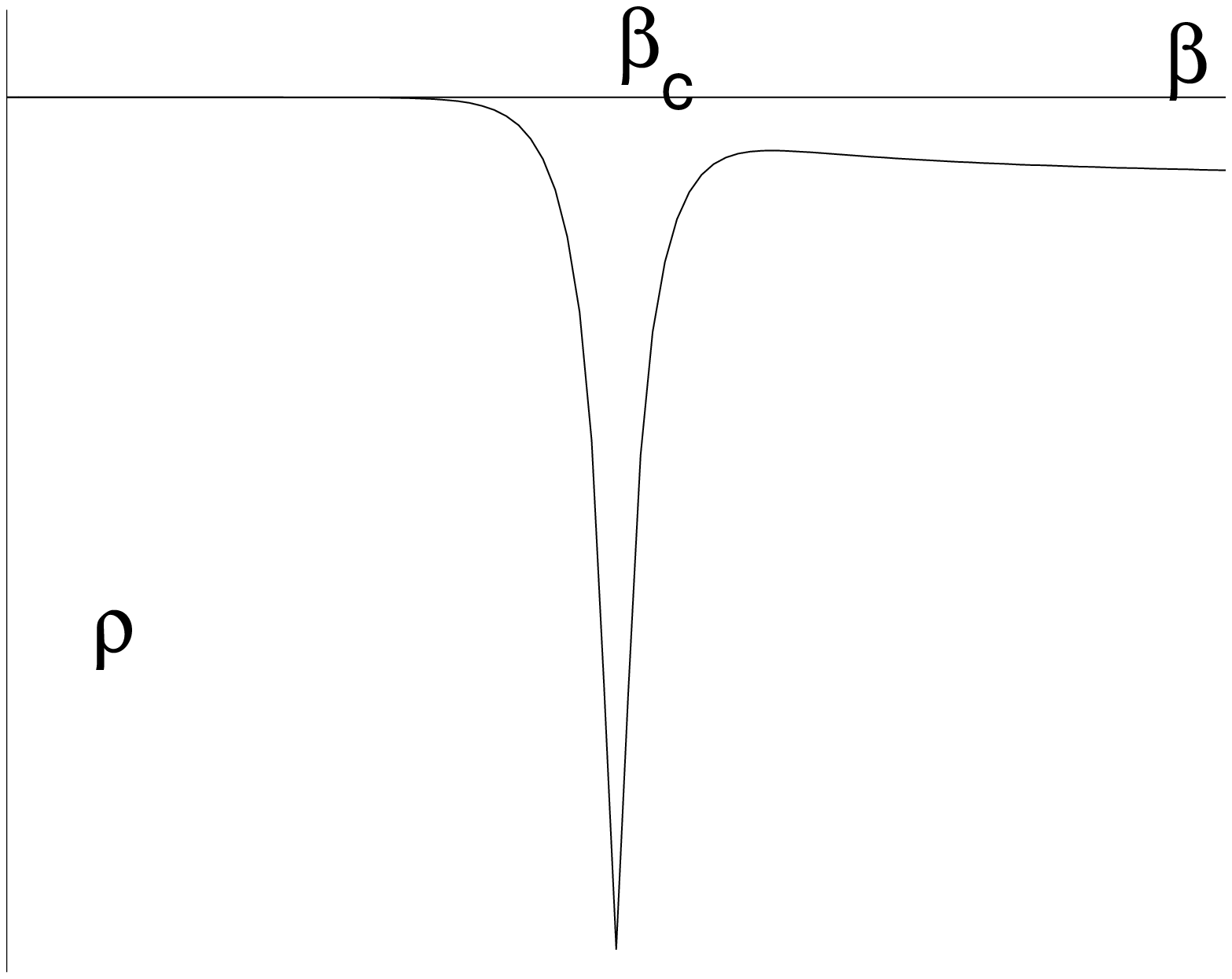}
}}
\caption{$\rho$ vs. $\beta$.}
\end{minipage}
\end{figure}

In the thermodynamical limit the shape of $\langle\mu\rangle$ becomes more and more
sharp at the transition point, or the negative peak in $\rho$ deeper and deeper.
$\rho$ goes to a finite limit for $T < T_c$, goes to $-\infty$ for $T > T_c$ as
$\rho \simeq - k L + k'$, ($k > 0$, $L$ lattice size), making $\langle\mu\rangle = 0$.
At the transition, where the correlation length goes large with respect the lattice
spacing, a scaling law is obeyed in pure gauge theory, of the form
\begin{equation}
\rho L^{-1/\nu} = f(\tau L^{1/\nu})\label{eq2}
\end{equation}
where $\tau = 1 - T/T_c$ and $\nu$ is the critical index of the correlation length.

Both $T_c$ and $\nu$ can be determined by fitting the scaling law to numerical
determinations on lattices with different spatial size $L$.

A number of investigations have been performed in the last year, to settle specific
points of this program. I will report on two of them.
\begin{itemize}
\item[1)]
How to reconcile the nonzero vev of a charged operator with gauge invariance. This
question is not very specific: it is also relevant for ordinary superconductors.
However it is worth of discussion, being systematically asked by different people. It
will be discussed in sect.2
\item[2)]
How does the construction work in the presence of dynamical quarks and how 
$\langle\mu\rangle$
is related to
the chiral order parameter $\langle\bar\psi \psi\rangle$.
\end{itemize}
\section{Gauge {Invariance$^8$}}
Superconductivity means condensation of charges in the ground state. The
relativistic version of the Ginzburg-Landau free energy is
\[ {\cal L} = -\frac{1}{4} F_{\mu\nu} F_{\mu\nu}
+ \frac{1}{2} (D_\mu\varphi0^* (D_\mu\varphi) +\frac{m^2}{2}
\varphi^*\varphi - \frac{\lambda}{4}(\varphi^*\varphi)^2\]
Density of free energy plays in statistical mechanics the same role as
effective lagrangean: it has to be used only at tree level.

$m^2$ and $\lambda$ depend on the temperature. If $m^2> 0$  the potential has the typical mexican hat shape and
$\varphi^*\varphi = \frac{m^2}{\lambda} \neq 0$ (Fig.3),
meaning that $\varphi$ itself is different from zero. 
\begin{figure}[t]
{\centerline{
\epsfxsize=0.75\textwidth %
\epsfbox{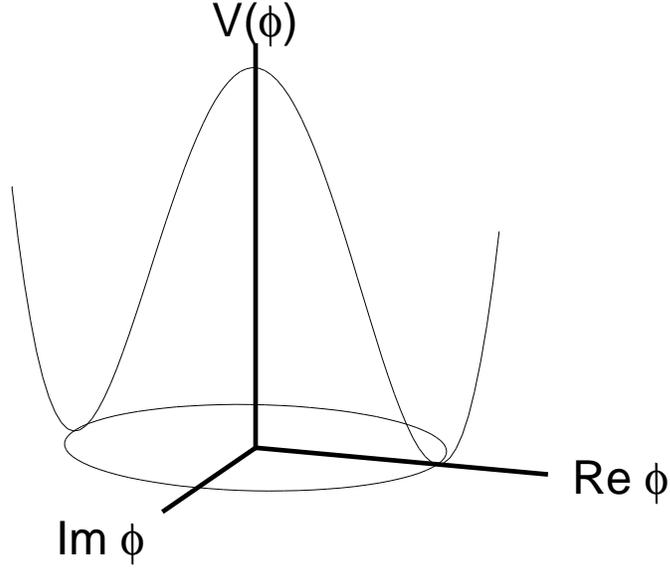} 
}}
\caption{$V(\phi) vs\phi$.}
\end{figure}
It can point,
however in any direction in the complex plane Re~$\varphi$ Im~$\varphi$
along the circle of minimum, (fig.3). It would seem that on the average
$\varphi = 0$ (Elitzur's theorem\cite{9}). However usually a gauge fixing is
made, e.g. the unitary gauge in which $\varphi$ is real and positive and then
$\varphi\neq 0$ is perfectly legitimate\cite{10}, it indicates that the ground state
is not an eigenstate of charge, but is a superposition of states with
different charges (Bogoliubov, Valatin).

In a gauge invariant formulation of the theory like lattice, however, 
the problem exists.
A microscopic
order parameter can be defined which is the vev of  a charged, gauge
invariant operator. Being gauge invariant it does not suffer of the
restrictions of Elitzur's theorem, and it signals at the same time that
vacuum is not $U(1)$ invariant.
The construction goes back to Dirac\cite{11}. Let $\varphi$ be any charged
operator. Under gauge transformations
\[ \varphi \to \varphi e^{i\theta}\]
Define
\begin{equation}
\tilde\varphi = \varphi \exp\left[ i \int d^4y C_\mu(x-y) A_\mu(y)
\right]
\label{eqg1}\end{equation}
with
\begin{equation}
\partial_\mu C_\mu = \delta^4(x-y)
\label{eqg2}\end{equation}
Then under a gauge transformation with parameter vanishing at
$x\to\infty$ the phase $\bar\theta$ of the second factor transforms as
\[\delta\bar\theta = 
 i \int d^4y C_\mu(x-y)\partial_\mu\Lambda =
-\int d^4 y \Lambda(y) \delta^4(x-y) = -\Lambda(x)\]
which cancels the phase change of $\varphi$. 

Under a global $U(1)$ $A_\mu$ is invariant and $\tilde \varphi$
transforms as any honest charged operator.

The additional phase factor in eq.(3) can be chosen to be  a
Schwinger string, but then the phase factor spoils cluster property of
$\tilde \varphi$. A choice which respects cluster property is Dirac choice\cite{11}
\begin{equation}
C_\mu(x) = (0, \vec C(\vec x-\vec y))\qquad
\vec C(\vec x-\vec y) =
\frac{1}{4\pi}\frac{\displaystyle \vec x-\vec y}{
\displaystyle |\vec x-\vec y|^3}
\label{eqg3}\end{equation}
which obeies eq.(1) and does not destroy cluster property.

Our operator $\mu$, used to detect dual superconductivity is magnetically
charged, but invariant under magnetic gauge transformations, like $\tilde
\varphi$, and Dirac type, eq.(\ref{eqg3}). So it provides by its vev a
perfectly legitimate disorder parameter.
\section{Full QCD}
It is  a usual statement that somehow the symmetry responsible for
confinement is different in pure gauge theories and in the presence of
quarks. In pure gauge the order parameter is the vev of Polyakov line, and
the symmetry $Z_N$, the centre of the group. Since in the presence of
quarks $Z_N$ is explicitely broken, people say that the true order
parameter is then the chiral condensate $\langle\bar\psi\psi\rangle$, which
signals spontaneous breaking of chiral symmetry, the pseudoscalar bosons
being the Goldstone particles. However also chiral symmetry is
explicitely broken by the electroweak quark masses.

Moreover, even if it is intuitively clear that confinement can produce
chiral
symmetry breaking, it is not clear what relation exists between chiral
symmetry and confinement. The numerical indication are that the two
transitions take place at the same temperature, but in principle
it is not understood why.

This state of the art is deeply unsatisfactory: if confinement has to be an
absolute property due to symmetry, to explain the very low upper limits put
by experiment on the rate of quark production in hadronic reactions, then
an exact symmetry should characterize the difference between confined and
unconfined phase.

Moreover there are many theoretical indications that the expansion in
$1/N_c$, $N_C$ being the number of colors, is a good expansion, and that
the theory at finite $N_c$ differs by small corrections from the limiting
theory at $N_C=\infty$.

If this is true the mechanism of confinement is fixed by the limiting
theory, and should be the same irrespective of $N_c$ and of the presence of
quarks, which contribute in non leading terms in the expansion.

If dual superconductivity in  all abelian projections is the symmetry
behind confinement, then it should equally work in full QCD.

A large scale program of simulations in on the way to verify that this is
the case.

Preliminary results show that indeed the symmetry pattern is the same in
full QCD as in quenched theory.

A complication to the analysis is that, while in pure gauge theory 
there is only
one length scale, say the physical correlation length, in the presence of
quarks, quark masses are present,and the limit of infinite volume is a two
scale problem. 

A systematic investigation is on the way on the APE100 computers in Pisa.

Of course the big theoretical open problem is still open, of identifying
the dual excitations which have non zero magnetic charge in all abelian
projections, and whose effective theory should hopefully give a weak coupling
description of the confined phase.

i thank all the people with whom i have collaborated in the research program,
in particular G. Paffuti.

\end{document}